# Analysis of Dual-beam Asymmetrical Torsional Bi-Material Cantilever for Temperature Sensing Applications

Yi Zheng

Department of Mechanical Engineering, Columbia University, New York, NY 10027 USA

Corresponding author: yz2308@columbia.edu

**ABSTRACT**

An extremely sensitive temperature measurement MEMS device is developed based on the principle of structural deflection in a bi-material cantilever caused by a difference in thermal expansion coefficients. A dual-beam asymmetrical geometry is used to produce a torsional response from the device. An analytical model is developed to predict the performance and optimize the free parameters of the device. In this work, it is performed to analyze the flexural and torsional eigenfrequencies as well as confirm the theoretical predictions of DC and AC response. Lastly, a procedure is developed to allow fabrication of the device using equipment available in the Columbia University clean room.

**INTRODUCTION**

It is desired to have a MEMS scale device which can detect very small changes in temperature [1]. One such method of achieving this is to construct a dual layer cantilever beam in which the two materials used have different values of thermal expansion. When the temperature of both materials is raised to a new value above the reference, the beam experiences a deflection that is similar to applying a load or moment at the end of the cantilever [2,21-24].

For this device, two of these beams are arranged in an asymmetrical configuration with one beam bending down and the other bending up. They are spaced some distance apart and are connected by a neutral piece. This configuration will produce a torsional response from the device rather than the normal flexural of an individual cantilever [3]. The benefits of this configuration over standard flexing are discussed in detail in the following section.

For this design, it is desired to produce a device which has beams that are 200 μm long in the axial direction and consists of a layer of Silicon Nitride (SiN) 500 nm thick coated with a layer of Gold (Au) 75 nm thick. These materials are chosen mostly due to their large difference in thermal expansion with $\alpha_{gold}$ being over 5 times that of $\alpha_{SiN}$. The width and spacing of the beams are not fixed and can be optimized for device performance [1,4].

Two geometries are considered for the project. One is in a "V" configuration, the other a "U" shape. These are shown in Figure 1. In the U geometry, the width of the end connecting piece is arbitrarily set to equal that of the cantilevers themselves. For the V geometry, the axial length of the beams is fixed so the size of the connecting piece is directly determined by the width and spacing (or relative angle) of the beams.

In this work, it is desired to construct an analytical model that can predict the device performance such as the relationship between temperature change and angular displacement (DC gain), torsional resonant frequency, and AC response. This can be done for varying values of beam width and spacing.

Additionally, an analysis will be performed on the device to confirm the theoretically calculated values and perform more in-depth eigenfrequency analysis [6-9].. Figures are produced showing the various modes of vibration and the angular deflection for selected temperatures, as shown in Fig. 1.

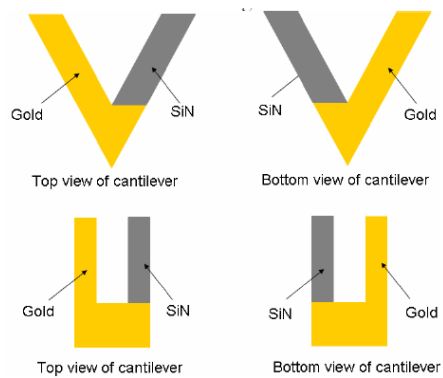

Figure 1: Profiles of device geometry

**ANALYTICAL MODEL**

In order to model the deflection angle as a function of change in temperature, the second order differential equation of torsional motion is used. This equation has the following form.

$$J_m \ddot{\phi} + b\dot{\phi} + k\phi = T$$

In this equation, $J_m$ is the mass moment of inertia of the object, b is the damping coefficient, k is the torsional spring constant of the device, T is the torque applied from a temperature change, and $\phi$ is the angle of deflection from horizontal on the end of the device.

To find the torque applied to the end of the device due to a temperature change, literature on bi-material cantilever beams was analyzed. This literature focused on single cantilever beams of two materials. One study modeled the deflection at the end of a single cantilever beam with an applied moment [5]. This study found the effective young's modulus of the beam, which was used in this model. Another study found the radius of curvature of a single bi-material beam as a function of change in temperature [7].

$$\frac{1}{r} = \frac{6w^2 E_1 E_2 t_1 t_2 (t_1 + t_2)(\alpha_1 - \alpha_2)\Delta T}{(wE_1 t_1^2)^2 + (wE_2 t_2^2)^2 + 2w^2 E_1 E_2 t_1 t_2 (2t_1^2 + 3t_1 t_2 + 2t_2^2)}$$

Taking the reciprocal of radius of curvature and multiplying by the square of the beam length and dividing by two gives the deflection at the tip of the beam. To approximate the force on the beam, standard beam deflection tables were consulted to find the point force at the tip of a single beam corresponding to the deflection found from literature [10-15].

$$F = \frac{6EI}{2L^3}\left(\frac{L^2}{2r}\right)$$

In this equation all of the properties are for a single cantilever beam, using the effective Young's Modulus found in literature [16-19]. The result of this equation is a force on the device as a function of change in temperature. Having a force on the device and knowing the distance from the axis of symmetry of its application, a torque on the device can be estimated as the force times the distance from the axis of symmetry. There are also two of these forces acting on the device, so the torque is doubled. In the case of the U geometry, this distance is half the gap width. In the case of the V geometry, the distance is estimated to be one third of the gap width.

The torsional spring constant is like a spring constant in that it represents an object's ability to resist a motion. This constant is simply the polar moment of inertia times the shear modulus divided by length for circular geometries. For rectangular geometries the relation is more involved [12].

$$k = \frac{G}{L}\left(wt^3\right)\left(\frac{1}{3} - 0.21\frac{t}{w}\left(1 - \frac{t^4}{12w^4}\right)\right)$$

This equation applies for rectangular geometries of width, w and thickness, t. For the geometries studied in this case, adjustments had to be made. The thickness and length used was that of the individual beams, and the shear modulus was calculated using the effective Young's Modulus and the Poisson's ratio of silicon nitride. For the U geometry, the k value of the gap was subtracted from the k of entire width of the device. For the V geometry, the k was calculated at the total width of the device, and the k values for the triangular gaps were subtracted. It was assumed that for triangular geometries, k is one half the k for rectangles of the same width and length.

To find the mass moment of inertia for the different geometries, different approaches were taken. For both geometries the effective masses of the beams were calculated and used as point masses at the centroid of the beams. For the V geometry, the end piece was also taken as two point masses at the centroid of the triangles. Mass moment of inertia for point masses is simply mass times the square of the distance from the rotational axis. The V geometry was therefore the sum of the mass moment of inertia of four point masses. There exists a formula for mass moment of inertia for rectangular geometries, and in the case of the U geometry, the polar moment of inertia was the sum of two point masses plus the polar moment of inertia of the end piece rectangle.

Damping occurs due to the effects of the air on the device. There is much literature on the subject of damping for single cantilever beams. A relation for damping at high resonant frequency for rectangular cross sections is used [12].

$$b = 2w\pi\mu\sqrt{\frac{\rho\omega_n}{2\mu}}$$

This equation applies to one beam. In this case μ is the fluid viscosity of air at room temperature, $\omega_n$ is the natural frequency of the device and is equal to the square root of the quantity k divided by $J_m$, ρ is the density of silicon nitride and w is the individual beam width. For each device the total b is two times the above equation as there are two beams.

Because some of these parameters were estimated, they are not entirely accurate. However, because they all scale linearly, correction factors can be added to make them match the analytical model. A table of the correction factors found to match data follows [20].

Table 1 Correction factors

| Parameter | U Geometry | V Geometry |
|---|---|---|
| $J_m$ | 0.10767 | – |
| k | 0.830132 | 0.364632 |
| F | 0.6323 | 2.2281 |

The reason these correction factors have been added is so that the optimization of the dimensions can be done accurately.

**POWER INPUT**

It is desired to be able to induce a temperature change in the device by applying an electric current through it. This ability is particularly useful for testing the device since any temperature is instantly achievable without having to alter the surrounding ambient conditions. Additionally, this may allow the device to function as a type of electro-thermal-mechanical actuator in which an applied current can generate a torque to be used for some other function.

Current is sent through one of the two fixed ends of the sensor and exits at the other which is attached to ground. The resistivity of the materials will cause energy to be dissipated internally in the device, raising its temperature. The rate at which heat is generated is given by:

$$P = IV = I^2 R.$$

Here, the electrical resistance, R is determined by: $R = \frac{\rho L}{A}$ where ρ=resistivity, L=length, and A=cross sectional area. The resistivity is a material property that has an almost linear dependence on temperature. The two material layers should be considered as two separate resistors connected in parallel with the total resistance given by:

$$R_T = \frac{R_1 \times R_2}{R_1 + R_2} \ (\Omega).$$

The rate at which energy is dissipated in the device must equal the rate at which heat is transferred to its surroundings. If this were not true, temperature would increase indefinitely and the device would melt. The heat transfer is modeled as a combination of conduction to the substrate and radiation to the surroundings with convection being ignored since it is often an order of magnitude less in MEMS devices.

Conduction is modeled under a 1D approximation for both materials in parallel as:

$$Q_{cond} = \frac{kA}{L}(T_2 - T_1).$$

Radiation is approximated as black-body with a complete view factor. The radiation equation which normally depends on T$^4$ can be linearized by introducing a radiation heat transfer coefficient[18].

$$Q_{rad} = h_{rad} A_{surf}(T_2 - T_1)$$

$$h_{rad} = \sigma(T_2^2 + T_1^2)(T_2 + T_1)$$

The input power and heat transfer are then set equal to one another so that:

$$I = \sqrt{\frac{Q_{cond} + Q_{rad}}{R}}.$$

The value of $T_1$ is set to an ambient temperature of 298K and through an iterative process, $T_2$, the device temperature can be solved for as a function of input current.

# FABRICATION

The designing process for the fabrication of the torsional bi-material cantilever temperature sensor was focused on fabricating the device at Columbia University. The Columbia University Clean Room is located at the CESPR building, where the Clean Room possesses state of the art fabrication equipments for creation and evaluation of nanoscale devices for the university [11,12].

The primary material the device composes of is gold; hence the Edward/BOC Thermal Evaporator will be used to thin-film deposit the gold to create the device. The other material required for the fabrication process will be silicon nitride, and that can be achieved in the clean room by chemical vapor deposition using the semi-core PECVD.

The process flow for fabrication of the device is described as follow:

Starting Material: p-type silicon substrate <100>

1. Clean
2. Photolithography for the alignment mask
3. Etch marks onto the SiO2
4. Strip Photoresist
5. Clean
6. Deposit a sacrificial layer by PECVD
7. Photolithography  (Au-1 Pattern)
8. Spin on Photoresist
9. Deposit Gold (Au-1) layer using thin-film deposition ~1800 C
10. Strip PR
11. Spin on PR
12. Photolithography (SiN-1 Pattern)
13. Deposit Silicon Nitride (SiN-1) layer using PECVD ~300 C
14. Strip PR

15. Spin on PR

16. Photolithography (Au-2 Pattern)

17. Deposit Gold (Au-2) layer using thin-film deposition ~1800 C

18. Strip PR

19. Spin on PR

20. Photolithography (Si-2 Pattern)

21. Deposit Silicon Nitride (SiN-2) layer using PECVD ~300 C

22. Strip PR

23. Test

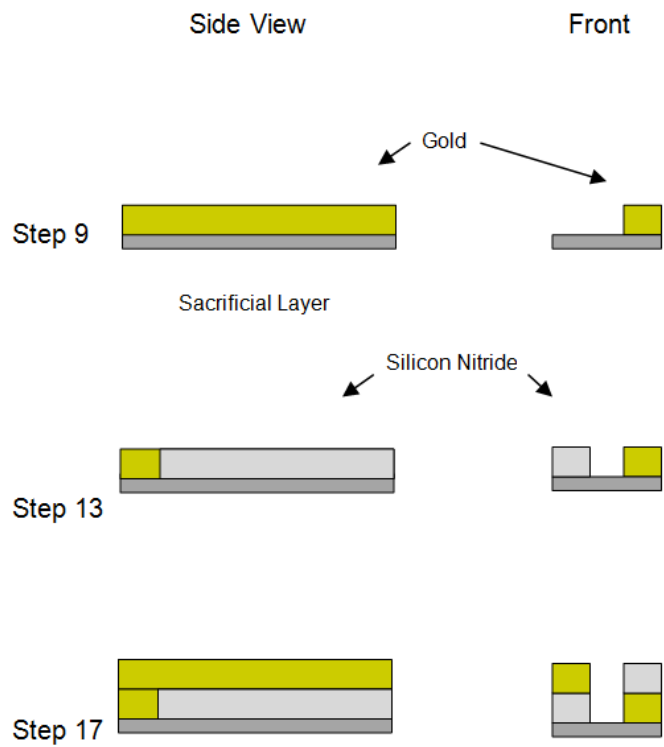

Figure 2: Fabrication of device

**RESIDUAL STRESS**

There is additional design possibilities beyond the ones mentioned previously. The two gold layers could be deposit using electrodeposition instead of thin-film deposition; the temperature of the operation would be lower comparatively to the thermal evaporation process and minimize residual stress. Additionally, another method to fabricate this cantilever system would be to deposit a thick gold layer, and back etch cavities on both sides of the gold film, and then followed by filling the cavities with Silicon Nitride. Finally to release the structure, the device can be released using DRIE.

## CONSLUSION

We developed an extremely sensitive torsional bi-material cantilever for temperature sensing applications, based on the principle of structural deflection caused by a difference in thermal expansion coefficients. In this paper, a dual-beam asymmetrical geometry is used to produce a torsional response from the device, for which an analytical model is developed to predict the performance and optimize the free parameters of the device. Lastly, a detailed procedure of the fabrication of the device is developed using equipment available in the Columbia University clean room.

## ACKNOWLEDGEMENT

The author would like to acknowledge the instruction with Professor Chee Wei Wong and discussions with Matthew Conwell, Ian McKinley and Xiaoyang Shi. This work was part of the graduate project (MECE E4212 Micro-electro-mechanical Systems) in Columbia University.